\documentclass{article}
\usepackage{spconf,amsmath,amssymb,graphicx,algorithm,booktabs,url,svg}
\usepackage{hyperref}
\usepackage[nolist,nohyperlinks]{acronym}

\title{SINGING LANGUAGE IDENTIFICATION USING A DEEP PHONOTACTIC APPROACH}
\name{Lenny Renault\textsuperscript{$\star$},
      Andrea Vaglio\textsuperscript{$\star \dagger$},
      Romain Hennequin\textsuperscript{$\star$}}
\address{
    \textsuperscript{$\star$}Deezer Research, \href{mailto:research@deezer.com}{\texttt{research@deezer.com}}\\
    \textsuperscript{$\dagger$}LTCI, Télécom Paris, Institut Polytechnique de Paris
}
\acrodef{SLID}{\emph{Singing Language Identification}}
\acrodef{ASR}{\emph{Automatic Speech Recognition}}
\acrodef{LID}{\emph{Language Identification}}
\acrodef{DNN}{\emph{Deep Neural Network}}
\acrodef{RNN}{\emph{Recurrent Neural Network}}
\acrodef{CRNN}{\emph{Convolutionnal Recurrent Network}}
\acrodef{LSTM}{\emph{Long Short-Term Memory}}
\acrodef{CTC}{\emph{Connectionist Temporal Classification}}
\acrodef{LRE}{\emph{Language Recognition Evaluation}}
\acrodef{UBM}{\emph{Universal Background Model}}
\acrodef{GMM}{\emph{Gaussian Model Mixtures}}
\acrodef{SVM}{\emph{Support Vector Machines}}
\acrodef{MFCC}{\emph{Mel Frequency Cepstral Coefficients}}
\acrodef{SAI}{\emph{Stabilized Auditory Images}}
\acrodef{TRAP}{\emph{Temporal Patterns}}
\acrodef{SDC}{\emph{Shifted Delta Cepstrum}}
\acrodef{IPA}{\emph{International Phonetic Alphabet}}

\begin{document}
%
\maketitle
\begin{abstract}

Extensive works have tackled \ac{LID} in the speech domain, however their application to the singing voice trails and performances on \ac{SLID} can be improved leveraging recent progresses made in other singing related tasks. This work presents a modernized phonotactic system for \ac{SLID} on polyphonic music: phoneme recognition is performed with a \ac{CTC}-based acoustic model trained with multilingual data, before language classification with a recurrent model based on the phonemes estimation. 
The full pipeline is trained and evaluated with a large and publicly available dataset, with unprecedented performances. First results of \ac{SLID} with out-of-set languages are also presented.

\end{abstract}
\section{Introduction}
\label{sec:intro}

Having semantically meaningful and accurate descriptions of songs is crucial for organizing and retrieving relevant tracks in a large musical catalog \cite{Turnbull2008_SemanticRetrieval}. Language tags are of particular interest for characterizing vocal music.
This information can easily be extracted using the song lyrics with a text-based language identifier \cite{Mahedero2005_NLP_lyrics}. However, lyrics are not ubiquitously available for consequent musical collections.
One then may want to estimate the song language from the frequently accessible metadata (e.g. song title, artist name). Yet, this method is limited as the metadata language can differ from the song language, and metadata may not contain enough information for retrieving the language \cite{Tsai2004}.
A more robust approach would be to extract the language information from the audio content. Such data is indeed always available, but the task is arguably more challenging.

The speech community has long tackled language recognition from audio data, notably with the \ac{LRE} series \cite{Martin2014NISTLR}. However, the task was scarcely transposed in the music domain and most of the techniques used for spoken \ac{LID} have yet to be adapted for \ac{SLID}.
The latter task is more difficult, as the prosodic specificities of languages are disturbed by the greater variabilities of the singing voice, in terms of pitch, pronunciation and vowels duration \cite{mesaros2010_automatic_reco_of_lyrics}. The musical accompaniment can be framed as noise and assumed to be loud and highly correlated with the signal of interest, being the voice.

Previous works on \ac{SLID} include acoustic-phonetic systems which characterize language-specific acoustic events and their distribution with carefully chosen acoustic features, such as \ac{MFCC} \cite{Schwenninger2006_lid_vocal_music}, \ac{SAI} \cite{Chandrasekhar2011_lid_video} and \ac{TRAP} \cite{Kruspe2014_GMM}. Statistical modeling and supervised classification are then applied to identify the language. In particular, Kruspe's system using the i-vector extraction technique obtains the current best performances on \ac{SLID} \cite{Kruspe2014_I_Vector}, with $78\%$ accuracy on a capella performances in $3$ languages.

Phonotactic approaches, on the other hand, try to identify phonemes from the audio and examine their combinations and sequences, which are distinctive from one language to another \cite{li_funda_to_practice}. These approaches are more resource-demanding as phoneme recognizers, or acoustic models, have to be trained.
Mehrabani et al. \cite{Mehrabani2011} use multiple language-specific phoneme recognizers trained on speech data, to then compute language likelihoods with n-gram models for each target language. While the performances are on par with Kruspe's i-vector-based approach \cite{Kruspe2014_I_Vector}, it is more complex to train and it is hardly scalable to a large set of languages.
In \cite{Kruspe2016_Phonotactic}, the author simplifies the approach by using a unique \ac{DNN}-based English phoneme recognizer and identifies the language from phoneme statistics. While singing data are included in the acoustic model training, the frame-wise phoneme annotations are obtained by a forced-alignment step, which leads to poorly annotated data. Also, this statistics-based language modeling overlooks the information contained in the phoneme transitions.

Recent works have trained new acoustic models with singing data and show great results in lyrics transcription \cite{Stoller2019_E2E_alignment}, lyrics-to-audio alignment \cite{gupta_2019, vaglio2020_multilingual} and explicit content detection \cite{vaglio2020_explicit}, using more recent \ac{DNN} techniques.
In this work, we propose to apply these advances to a phonotactic \ac{SLID} system: in particular, the usage of the \ac{CTC} algorithm allows the acoustic model to be trained with DALI, a large multilingual singing dataset \cite{Meseguer-Brocal2018}, while alleviating the need for frame-level aligned lyrics.
For language modeling, we use a recurrent architecture that can capture temporal information in phoneme estimation sequences.
We show that our system outperforms the previous state-of-the-art 
in a standard closed-set scenario, obtaining a $91.7\%$ balanced accuracy score on polyphonic songs in $5$ languages. We also investigate a harder setup with out-of-set languages where we can acknowledge the limits of our model.
In Section \ref{sec:proposed_system}, we describe the key aspects of our deep phonotactic \ac{SLID} approach.
The dataset, baselines and implementation details are given in Section \ref{sec:experimental_setup}.
Results are presented in Section \ref{sec:results}, providing a first reproducible benchmark on the public DALI dataset\footnote{The dataset split can be found at \url{https://github.com/deezer/SingingLanguageIdentification}}.

\section{Proposed system}
\label{sec:proposed_system}

 \begin{figure} 
  \centering
  \includegraphics[width=\columnwidth]{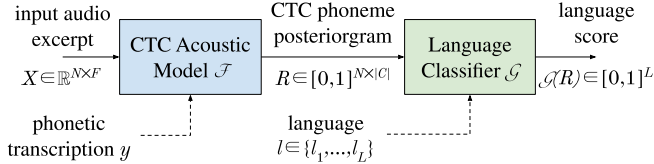}
  \caption{Overview of the proposed \ac{SLID} system. During training, the phonetic transcription of the lyrics $y$ and the corresponding language label $l$ of the excerpt $X$ are provided.}
  \label{fig:system_overview}
\end{figure}

As in previous \ac{SLID} works, we frame the problem as a multiclass classification task. The system takes as input audio features of a musical excerpt $X \in \mathbb{R}^{N \times F}$, with $N$ the number of time frames and $F$ the feature dimensionality. The system should then estimate the language $l$ used in the musical excerpt, among a set of $L$ languages $\{l_1, l_2, ..., l_L\}$.

Our deep phonotactic system, as illustrated by Figure \ref{fig:system_overview}, is composed of two main models: an acoustic model $\mathcal{F}$ for phoneme estimation, followed by a language classifier $\mathcal{G}$. The acoustic model estimates the occurring phonemes in the input audio $X$ by producing a $|C|$-dimensional vector of probabilities at each time frame, with $C$ the set of characters supported by $\mathcal{F}$. Here, $C$ encompasses the \ac{IPA} symbols appearing in the training excerpts, a word-boundary ``space" token, an instrumental ``I" token and the blank token $\epsilon$ introduced by the \ac{CTC} algorithm in Section \ref{sec:phoneme_recognition}. The sequence of phoneme probability vectors is referred as the posteriorgram $R := \mathcal{F}(X) \in [0,1]^{N \times |C|}$.

The language classifier $\mathcal{G}$ then produces a language probability vector score $\mathcal{G}(R) \in [0,1]^{L}$ from the posteriorgram $R$. The language decision is finally taken from this vector score:
\begin{equation}
    \hat{l} = \operatorname*{arg\,max}_{l \in \{l_1, l_2,...,l_L\}} 
    [\mathcal{G}(\mathcal{F}(X))]_l .
\end{equation}

Previous phonotactic approaches on \ac{SLID} made distinct training of the two models, with different training sets \cite{Mehrabani2011,Kruspe2016_Phonotactic}. $\mathcal{F}$ needs the phonetic transcription $y$ of each training excerpt $X$, whereas $\mathcal{G}$ needs the posteriorgram representation $R$ and the language label $l$ of its training excerpts. We use the same dataset of musical excerpts for both model training.  Following the works on joint \ac{ASR} and \ac{LID} \cite{watanabe_joint_asr_lid}, we train both models simultaneously. The joint loss function can be expressed as:
\begin{equation}
    \mathcal{L}_{Joint}(R, \hat{l}, y, l) = \mathcal{L}_{\text{CTC}}(R, y) + \lambda \mathcal{L}_{\text{LID}}(\hat{l}, l) ,
\end{equation}
with $\lambda$ the weight given to the cross-entropy \ac{LID} loss with regard to the \ac{CTC} loss.
The balance between the two losses is decisive for the system performance: various strategies are considered and described in Section \ref{sec:training_strategies}.

\subsection{Phoneme recognition}
\label{sec:phoneme_recognition}

For the acoustic model $\mathcal{F}$, we rely on a \ac{CRNN} trained with the \ac{CTC} algorithm, as in \cite{vaglio2020_explicit}. \ac{CTC}-based acoustic models were successfully implemented for singing-related tasks, such as lyrics-to-audio alignment \cite{Stoller2019_E2E_alignment, vaglio2020_multilingual} and keyword spotting \cite{vaglio2020_explicit}. Following their works, we also employ a singing voice separation pre-processing step during training and inference, which improves performance over using polyphonic data in \cite{Stoller2019_E2E_alignment}. For the recurrent layers, we choose bidirectional \ac{LSTM} cells to take the full sequence into account when predicting characters at each time frame.

The \ac{CTC} algorithm enables to train a \ac{RNN} with weakly aligned data, e.g. at word or line level, by introducing a blank token $\epsilon$ in the set $C$ of characters supported by the model. The associated loss function computes the probability of an output sequence by marginalizing over all possible alignments with the input. Following the work in \cite{vaglio2020_multilingual}, the output sequences are composed of multilingual phonemes according to the \ac{IPA}. As the \ac{CTC} loss function is differentiable, network training can be done with any gradient descent algorithm, by providing the phonetic transcription $y$ of the segment lyrics. Further details on the \ac{CTC} algorithm can be found in \cite{Graves_2006_CTC}.

\subsection{Language Classifier}

The language classifier $\mathcal{G}$ is built upon a \ac{RNN}. The usage of recurrent architectures has been successful for end-to-end spoken \ac{LID} \cite{Trong2016_E2E, Cai2019_UtterancelevelE2E_LID_CNNBLSTM}. Here, the phoneme posteriorgram representation is given as input, instead of the raw acoustic features extracted from the audio excerpt. Bidirectional \ac{LSTM} layers are chosen with the last layer only outputting a single probability vector for the whole input segment. This architecture takes the combination of phonemes into account, as in n-gram modeling \cite{Mehrabani2011}, but with confidence scores on the phoneme predictions by using the full probability vectors, as in statistical modeling \cite{Kruspe2016_Phonotactic}. To avoid vanishing gradient on very long input sequence \cite{hochreiter2001gradient}, we choose to perform \ac{SLID} on fixed-length segments for a given song. Song language is then inferred from the mean of language scores output by the system on all segments.

\section{Experimental setup}
\label{sec:experimental_setup}

\subsection{Dataset}
All versions of our system are trained and tested on tracks from the DALI dataset \cite{Meseguer-Brocal2018}. This dataset contains $5358$ songs of various western genres with the lyrics annotations at word-level and song-level language labels. All tracks are downsampled to 16kHz and converted to mono. Musical accompaniments are removed by vocal extraction with Spleeter \cite{spleeter}.

We design two language sets from this dataset: a \emph{closed-set} scenario and an \emph{open-set} scenario.
The closed-set retains languages with more than $10$ hours of data each: English, French, German, Italian and Spanish. The open-set also adds a sixth label ``Others" regrouping low-resource languages (Dutch, Finnish, Portuguese, Polish). Train, validation and test sets are obtained with a $80\%$-$10\%$-$10\%$ language-wise and artist-aware split \cite{Flexer2007_artist_filtering}.
Songs in neither target nor low-resource languages are also labelled as ``Others" and added to the open-set test set only. These out-of-domain samples help monitoring the generalization of out-of-set modeling learned from the subset of in-domain ``Others" languages.
As English is over-represented in the dataset, all systems and baselines are trained with a class-weighted \ac{LID} objective function.

Our system performs \ac{SLID} at segment-level. Each song is split into $20$s segments with a $0.5$ overlapping factor between two consecutive segments. Segment lyrics are retrieved using the word-level annotations from DALI and decomposed into \ac{IPA} symbols using Phonemizer \cite{phonemizer}.
Collecting all phonemes occurring in the training segments and adding the space, instrumental and blank tokens, the total number of characters obtained is $|C|=66$ in the closed-set scenario and $|C|=71$ in the open-set scenario.
For the segment language label, the FastText language identifier \cite{joulin2016fasttext} is used on the segment lyrics. A segment is labeled \emph{instrumental} when it has less than $3$ words, 
or \emph{ambiguous} when the lyrics repetitiveness or the FastText non-confidence score is above an empirically found threshold.
During inference, \emph{ambiguous} and \emph{instrumental} scores are not taken into account when estimating the song language from segment language scores.

\subsection{Baseline systems}
Two baseline systems are implemented for comparison with our system. The \emph{Metadata} baseline is a text-based language identifier using the artist name and song title metadata provided with the DALI dataset. The language is extracted using the FastText language identifier \cite{joulin2016fasttext}.

The \emph{i-vector} baseline is an acoustic-phonetic i-vector-based system, as in \cite{Kruspe2014_I_Vector}. Implemented with the Kaldi \ac{LRE} receipt \cite{kaldi}, this system computes $600$-dimensional i-vector per vocal-isolated song. Sequences of $20$ \ac{MFCC} feature vectors are extracted then modeled by a \ac{GMM}-based \ac{UBM}. Supervised language classification is performed from the i-vector representation of the song using \ac{SVM} with a cosine kernel, as in \cite{Kruspe2014_I_Vector, dehak_ivec}.

\subsection{Acoustic model architecture}
\label{sec:Acoustic_model_implementation}

40 Mel-scale log filterbanks coefficients and energy features, plus deltas and double-deltas are computed from the extracted vocals using a $32$ms Hann window with $0.5$ overlap. The input feature sequences are downsampled by two sub-modules each composed of a 2D-convolutional layer (32 filters with kernel size $3\times3$), a ReLU activation function and a $2\times3$ max-pooling layer: sequence length is thus divided by $4$. 

The recurrent part of the acoustic model is composed of 3 bidirectional \ac{LSTM} layers with $256$-dimensional hidden states. Dropout and recurrent dropout of $0.1$ each is applied. Finally a time-distributed dense layer and a softmax activation function are applied for obtaining per-frame character probability vectors from $C$. The \ac{CTC} layer and objective function implementations are taken from \cite{ctcmodel_implementation}.

\subsection{Language classifier}
\label{sec:language_classifier_implementation}

Inputted posteriorgrams are pre-processed by a deterministic cleaning module: frames with $\epsilon$-emission probability $p(\epsilon) > 95\%$ are removed, to account only for frames with actual phoneme predictions.

The language classifier model is composed of $2$ bidirectionnal \ac{LSTM} layers with $64$-dimensional hidden states each. The second layer outputs a single vector per segment, which is processed by a dense layer with a softmax activation function to produce one language probability vector. Recurrent layers have a $0.1$ recurrent dropout factor and $0.2$ dropout is applied between each layer. A class-weighted categorical cross-entropy loss function is used for training the model given the one-hot encoded language labels.

\subsection{Training strategies for our approach}
\label{sec:training_strategies}

We test two strategies for training our system, implemented in Tensorflow. Each training variant relies on the ADAM optimization algorithm \cite{Kingma2015_Adam} with a learning rate of $10^{-3}$, a batching size of $32$ and validation-based early stopping.

The \emph{2-step} variant first trains the acoustic model $\mathcal{F}$ alone.
The language classifier $\mathcal{G}$ is then trained for \ac{SLID} from the posteriorgrams of the training segments computed by $\mathcal{F}$.
The \emph{Joint} variant trains both models at the same time from scratch. With hyper-parameter tuning, we found that training the system with a loss balance $\lambda=0.1$, then fine-tuning it with $\lambda=100$ yields the best performances on the validation set.



\subsection{Ablation study}
We evaluate the relevance of our system parts by designing two simplified systems for comparison.
The \emph{E2E} system is an end-to-end approach to \ac{SLID} with the same architecture as the \emph{Joint} variant, except for the \ac{CTC} component which is removed from the loss function. The phoneme recognition task is ignored as the model is solely trained to identify the language in song segments.

The \emph{Statistics} system is a modified \emph{2-step} variant. Instead of the recurrent layers, the language classifier is a pooling step of the mean and variance statistics of each phoneme class over the full song length. Song language is directly predicted from these statistic vectors using \ac{SVM}. This system is analogous to a modernized version of \cite{Kruspe2016_Phonotactic}, with a \ac{CTC}-based acoustic model instead of the \ac{DNN}-based one.

\section{Results}
\label{sec:results}

\subsection{Performances in the closed-set scenario}
The results of the evaluation of our systems on the test songs in a closed-set scenario are reported Table \ref{tab:perf_CS}.

\begin{table}[H]
\small
\centering
\begin{tabular}{@{}rcc@{}}
\toprule
System     & bAccuracy (\%)  & F1-score (\%) \\ \midrule 
Metadata   & 76.48 (3.98)    & 76.71 (3.45)  \\ 
i-vector   & 77.26 (3.88)    & 67.78 (3.57)  \\ 
E2E        & 59.90 (4.33)    & 65.43 (4.47)  \\ 
Statistics & 88.46 (3.04)    & 89.00 (2.95)  \\ 
2-step     & 88.62 (3.03)    & 90.75 (2.62)  \\ 
Joint      & \textbf{91.74 (2.70)} & \textbf{92.39 (2.31)} \\ \bottomrule 
\end{tabular}%
\caption{Systems evaluation in the closed-set scenario. Measured by balanced accuracy (bAccuracy) and macro-averaged F1-score (with standard errors in parenthesis).}
\label{tab:perf_CS}
\end{table}

All phonotactic approaches (\emph{Statistics}, \emph{2-step} and \emph{Joint}) outperform the \emph{Metadata} baseline, on the contrary of the \emph{E2E} system. The phonetic information contained in the audio data is thus better suited for estimating the language than common metadata. Reliable estimations from the raw audio can not be achieved with a naive end-to-end approach and seems to require more refined techniques.
Our deep phonotactic system also significantly outperforms the re-implemented state-of-the-art \emph{i-vector} system.
In particular, joint training of the acoustic model and language classifier further improves the system performance, as the \emph{Joint} variant yields the best overall scores, with $91.7\%$ of balanced accuracy.

Regarding the efficiency of each system part, the \emph{Statistics} system has better performances that the \emph{i-vector} baseline, which was not the case between the two analog approaches from Kruspe \cite{Kruspe2014_I_Vector, Kruspe2016_Phonotactic}. Hence, our \ac{CTC}-based acoustic model seems to offer better modeling capability than the \ac{DNN}-based model from \cite{Kruspe2016_Phonotactic}.
The \emph{2-step} variant does not significantly outperform the \emph{Statistics} system, which implies that the language classifier can be improved.
Finally, even though the side phoneme recognition task requires more detailed information for training, it proves to be profitable for \ac{SLID} since the \emph{2-step} and \emph{Joint} systems outperform the \emph{E2E} baseline.

\subsection{Performances in the open-set scenario}
\label{sec:performances_open_set}

\begin{table}
\centering
\resizebox{\linewidth}{!}{%
\begin{tabular}{@{}rcc|cc@{}}
\toprule
System     & bAccuracy (\%) & F1-score (\%) & Target (\%) & Others (\%) \\ \midrule
Metadata   & 70.16 (3.46) & 70.52 (3.08) & 73.30 (3.45) & 56.60 (4.73) \\
i-vector   & 70.87 (3.16) & 54.79 (2.90) & 58.74 (3.18) & 35.06 (4.92) \\
E2E        & 39.57 (3.17) & 35.78 (2.57) & 42.93 (3.07) & 0.00 (0.00) \\
Statistics & \textbf{83.30 (2.83)} & \textbf{80.28 (2.79)} & \textbf{81.70 (3.03))} & \textbf{73.14 (3.79)} \\
2-step     & 78.49 (2.77) & 74.35 (2.92) & 79.89 (3.14) & 46.62 (5.40) \\
Joint      & 72.89 (2.86) & 64.46 (3.14) & 72.02 (3.51) & 26.67 (5.35) \\ \bottomrule
\end{tabular}%
}
\caption{Systems evaluation in the open-set scenario. Measured by balanced accuracy (bAccuracy) and macro-averaged F1-score. Macro-averaged F1-score on the target languages and the F1-score on the ``Others" class are also presented. Standard errors are in parenthesis.}
\label{tab:perf_OS_song}
\end{table}

The results of the evaluation of our systems on the test songs in the open-set scenario are reported Table \ref{tab:perf_OS_song}.
All phonotactic systems still outperform the \emph{Metadata}, \emph{E2E} and \emph{i-vector} approaches.
However, both variants of our deep phonotactic system are less robust to the introduction of the ``Others" class than the simpler \emph{Statistics} system. Indeed, they seem to overfit on the ``Others" training data.
It can be explained as this class has a greater linguistic variability than other classes but has the same amount of data as a low-resource target language.
This effect is further demonstrated in Table \ref{tab:perf_unseen_vs_seen} as only the \emph{Statistics} system can generalize the out-of-set modeling to out-of-domain languages unseen during training.


\begin{table}[H]
\centering
\resizebox{\linewidth}{!}{%
\begin{tabular}{@{}rcc@{}}
\toprule
System     & In-domain ``Others" (\%) & Out-of-domain ``Others" (\%)\\ \midrule
i-vector   & 50.00 (10.66)            & 20.00 (4.46) \\\
E2E        & 0.00 (0.00)              & 0.00 (0.00) \\
Statistics & \textbf{86.36 (7.29)}    & \textbf{56.25 (5.58)} \\\
2-step     & 63.64 (10.28)            & 21.25 (4.61) \\
Joint      & 31.82 (9.87)            & 11.25 (3.56) \\ \bottomrule
\end{tabular}%
}
\caption{Performances comparison on ``Others" labelled test songs in in-domain and out-of-domain languages cases. Measured by accuracy (with standard errors in parenthesis).}
\label{tab:perf_unseen_vs_seen}
\end{table}

\section{Conclusion}
\label{sec:conclusion}

We investigate modernized phonotactic systems for \ac{SLID} on polyphonic music, using recurrent models for both phoneme recognition and language classification. Trained on a publicly available multilingual dataset, the proposed system outperforms metadata-based and the previous state-of-the-art \ac{SLID} approaches.
The \ac{CTC}-based acoustic model greatly contributes to the performance increase, both in closed-set and open-set scenarios. However, the proposed language classifier hardly exceeds statistical modeling in a closed-set scenario, and deteriorates with out-of-set languages.
Future works would focus on exploring hierarchical language modeling techniques for \ac{SLID} with out-of-set languages, taking inspiration from the speech literature \cite{trong2018_staircase}.


\vfill\pagebreak

\bibliographystyle{IEEEbib}
{\footnotesize
\bibliography{strings,refs}}

\end{document}